\documentstyle[preprint,aps,epsf]{revtex}

\begin{document}

\title{Thermopower in p-type GaAs/AlGaAs layers}

\author{M. Tsaousidou, P. N. Butcher}

\address{Department of Physics, University of Warwick, Coventry CV4 7AL,
England}

\author{P. A. Crump, R. J. Hyndman, B. L. Gallagher}

\address{Department of Physics, University of Nottingham, NG7 2RD, England}

\maketitle
\begin {abstract}
We explain for the first time thermopower data for p-type
GaAs/AlGaAs layers. The data span a temperature range 0.2
K$\leq$T$\leq$1.2 K. We calculate both the diffusion $S^{d}$ and the
phonon-drag $S^{g}$ contributions to the thermopower. We find that
$S^{d}$ is significant for temperatures up to $\sim$ 0.3 K while at
higher temperatures $S^{g}$ dominates. The calculated and measured
$S^{g}$ agree very well with {\bf no} adjustable parameters.
\end{abstract}

\section{Introduction}
We investigate the thermopower $S$ of two-dimensional hole gases
(2DHGs) in p-type GaAs/AlGaAs. $S$ is defined by the relation ${\bf
E}=S\nabla T$, where ${\bf E}$ is the electric field, $\nabla T$ is
the temperature gradient and the current density ${\bf J}$ is zero.
We examine the diffusion, $S^{d}$, and the phonon-drag, $S^{g}$,
contributions to $S$. $S^{d}$ is due to the hole diffusion along
$\nabla T$ while $S^{g}$ is due to the momentum transfer from 3D
acoustic phonons through the hole-phonon coupling. We examine two
samples with hole densities $P_{s}$: 0.93 10$^{15}$m$^{-2}$ and 1.25
10$^{15}$m$^{-2}$ and mobilities close to 30 m$^{2}$/Vs. Hereafter
we refer to the sample with the lower $P_{s}$ as Sa1 and the one
with the higher $P_{s}$ as Sa2. The 2D hole layers are grown in the
(311) direction. The complexity of the valence band in p-type
GaAs/AlGaAs \cite{Go,Br,Ek} makes the calculation of $S$ a difficult
task. However, because the 2DHGs have low $P_{s}$ we can make the
following assumptions to simplify our analysis. We assume that only
the top heavy hole subband is occupied, that it is isotropic and
spin splitting is not important \cite{Go}. The value of the in-plane
effective mass $m^{*}$ is $0.3\,m_{e}$ which is in good agreement
with the experiment\cite{St}. The Fermi temperatures $T_{F}$ for Sa1
and Sa2 are 8.6~K and 11.6~K, and consequently for
0.2~K~$\leq$~T~$\leq$~1.2~K the 2DHGs are degenerate.

\section{Theory}
The diffusion contribution to the thermopower $S$ for a degenerate 2DHG is
given by \cite{Zi,Fl1}:
\begin{equation}
S^{d}=(p+1)\frac{\pi^{2}k_{B}^{2}T}{3|e|E_{F}},
\end{equation}
where $p=(\partial ln\tau/\partial lnE)_{E_{F}}$ with $\tau$ being the
hole transport lifetime
and $E$ the hole energy (with $E_{F}$ the Fermi energy).
The value of $p$ depends
on the type of scattering.

The phonon-drag contribution $S^{g}$ is calculated
by using a refined formula which is based
on Cantrell and Butcher's paper \cite{Ca} and includes
non-degeneracy effects and screening of the hole-phonon
interaction\cite{Zi,Fl1}:
\begin{equation}
S^{g}=\frac{2\tau_{F}}{\sigma}\sum_{m}\int_{0}^{\infty}\int_{-\infty}^{\infty}
dq\,dq_{z}\,q^{2}\,C({{\bf
Q}}) \int_{0}^{\infty}
du f^{0}(u^{2}+\gamma)\,[1-f^{0}(u^{2}+\gamma+\hbar\omega_{{\bf
Q}})],
\end{equation}
where ${\tau_{F}}$ is the hole transport lifetime at $E_{F}$,
$\sigma$ is the hole conductivity, $f^{0}$ is
the Fermi-Dirac distribution function, and $\omega_{{\bf Q}}$ is the frequency
of phonons with wave vector ${\bf Q}=({\bf q},q_{z})$ and mode $m$.
Moreover:
$u^{2}=E-\gamma$ where the expression of $\gamma$ is given in reference 5.
Finally, $C({\bf Q})$ is:
\begin{equation}
C({\bf Q})=\frac{|e|(2m^{*})^{1/2}l_{p}\Xi^{2}({\bf Q})N^{0}({\bf
Q})Q|Z_{11}(q_{z})|^{2}}{16\pi^{3}\hbar
k_{B}T^{2}\rho\epsilon^{2}(q)},
\end{equation}
where $l_{p}$ is the phonon mean free path,
$N^{0}$ is the Bose-Einstein distribution function, $\rho$ is the density
of the material, $\epsilon(q)$ is the dielectric function given by Gold and
Dolgopolov \cite{Gold}, and
$|Z_{11}(q_{z})|^{2}$ accounts for the finite thickness of the 2DHG
\cite{Zi}. Finally,
$\Xi({\bf
Q})$ is the
`effective' acoustic potential describing the hole-phonon coupling.
In GaAs $\Xi^{2}({\bf Q})$
accounts for both deformation potential and piezoelectric
coupling. For the longitudinal branch and for each of the transverse branches
$\Xi^{2}({\bf Q})$ is \cite{Zi}: $\Xi^{2}_{d}+[(eh_{14})^{2}A_{l}/q^{2}]$ and
$(eh_{14})^{2}A_{t}/q^{2}$, respectively. Here, $\Xi_{d}$ is the
deformation potential constant, $h_{14}$ is the piezoelectric constant, and
$A_{l}$, $A_{t}$ are the anisotropy factors given in reference 5.

\section{Results and Discussion}
Figures 1(a) and 1(b) show the measured (circles) and the calculated
(solid lines) values of S in Sa1 and Sa2, respectively. The straight
lines correspond to $S^{d}$ and the dashed lines to $S^{g}$. We see
that $S^{g}$ becomes dominant for $T>0.3$~K. $S^{d}$ is fitted by
the lines 40T $\mu V/K$ for Sa1 and 30T $\mu V/K$ for Sa2. Then from
Eq.~(1) we calculate $p=0.2$ for both Sa1 and Sa2. The small value
of $p$ implies a weak energy dependence of $\tau$ which is in good
agreement with mobility data in 2DHGs similar to those we examine
here \cite{He}.

$S^{g}$ is calculated by using Eq.~(2). The phonon mean free paths
$l_{p}$ are determined from the measured thermal conductivity
$\lambda$ (not shown here) by using the formula \cite{Sm}:
$l_{p}=(\lambda/T^{3})15\hbar^{3}{\bar
v}_{s}^{2}/2\pi^{2}k_{B}^{4}$, where ${\bar v}_{s}^{-3}$ is the
average inverse cube speed of sound for the three acoustic modes in
GaAs. The data for $\lambda$ show a $T^{3}$ dependence and $l_{p}$
is 1.6 mm for Sa1 and 1.4 mm for Sa2. The values of the other
material parameters used here are the standard ones for GaAs
\cite{Zi}. The usual unknown in the $S^{g}$ calculations is
$\Xi_{d}$. Here we use $\Xi_{d}=12.5$ eV. However, we find that the
piezoelectric contribution to $S^{g}$ is dominant over the
deformation potential one at all the temperatures considered. The
former is $\sim 60\%$ of the total $S^{g}$ at $T=1.2$ K and
increases to $\sim 95\%$ at $T=0.2$ K. Consequently, the uncertainty
in the value of $\Xi_{d}$ does not effect our results and,
practically speaking, the calculation of $S^{g}$ does not involve
any adjustable parameters. The good agreement between the
theoretical and experimental values of $S^{g}$ gives us confidence
in the validity of our theoretical considerations and suggests the
extension of the experiments and the calculations of $S^{g}$ to
higher temperatures where the deformation potential coupling becomes
important and $\Xi_{d}$ can be determined from the data.

Finally, Figures 2(a) and 2(b) depict the ratio $S/\lambda$ for Sa1
and Sa2, respectively. The circles are the experimental values and
the crosses the theoretical predictions. At $T<0.3$~K where $S^{d}$
dominates we expect $S/\lambda$ to vary as $T^{-2}$ which is
observed in Figure 2. At higher T both experiment and theory show a
maximum in $S/\lambda$. This is a Kohn resonance occurring when the
dominant phonon wave vector \cite{Sm} ${\bar q}=5k_{B}T/\hbar {\bar
v}_{s}$ becomes equal to  $2 k_{F}$, where $k_{F}$ is the Fermi wave
number. The temperatures at which we expect the maximum are 0.8 K
for Sa1 and 0.9 K for Sa2 which are in good agreement with the data.

In conclusion, we have calculated $S$ in two p-type
GaAs/AlGaAs layers and found a good agreement with the experiment.
$S^{g}$ is dominant for $T>0.3$~K. We suggest an
extension of the experiments to $T>1$ K where comparison with the
theory can determine the deformation potential constant $\Xi_{d}$.

\section*{Acknowledgements}
M.T. and P.N.B. wish to acknowledge the United Kingdom Engineering
and Physical Sciences Research Committee for financial support.

\begin{figure}
\epsfysize=7.5in
\centering
\epsffile{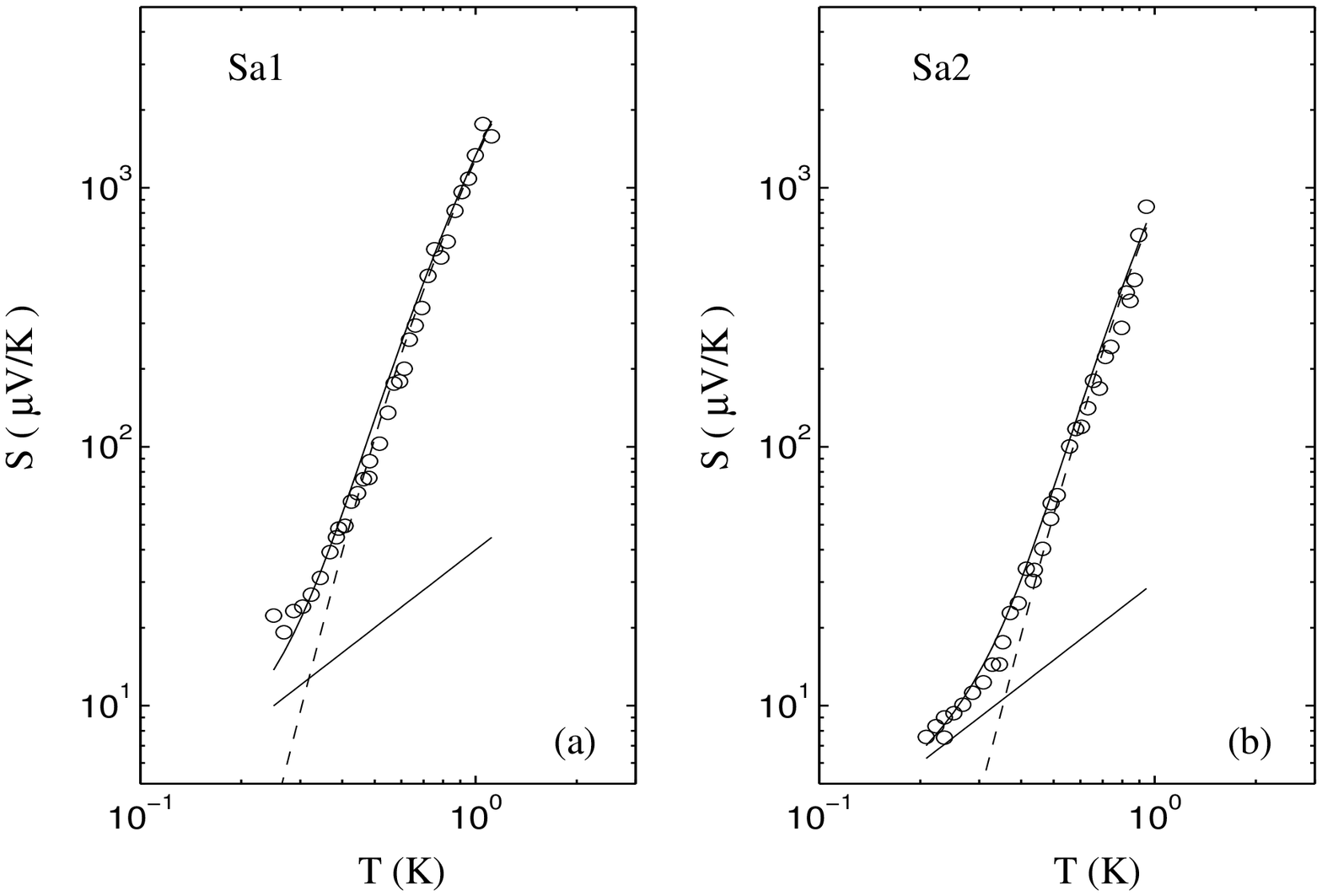}
\caption{Thermopower $S$ as a
function of T at: a) $P_{s}=0.93\times 10^{15}$m$^{-2}$ and b)
$P_{s}=1.25\times 10^{15}$m$^{-2}$. The circles are the experimental
data, and the solid line is the theoretical prediction. $S^{d}$ and
$S^{g}$ are shown by the straight and dashed lines, respectively.}
\end{figure}

\begin{figure}
\epsfysize=8in
\centering
\epsffile{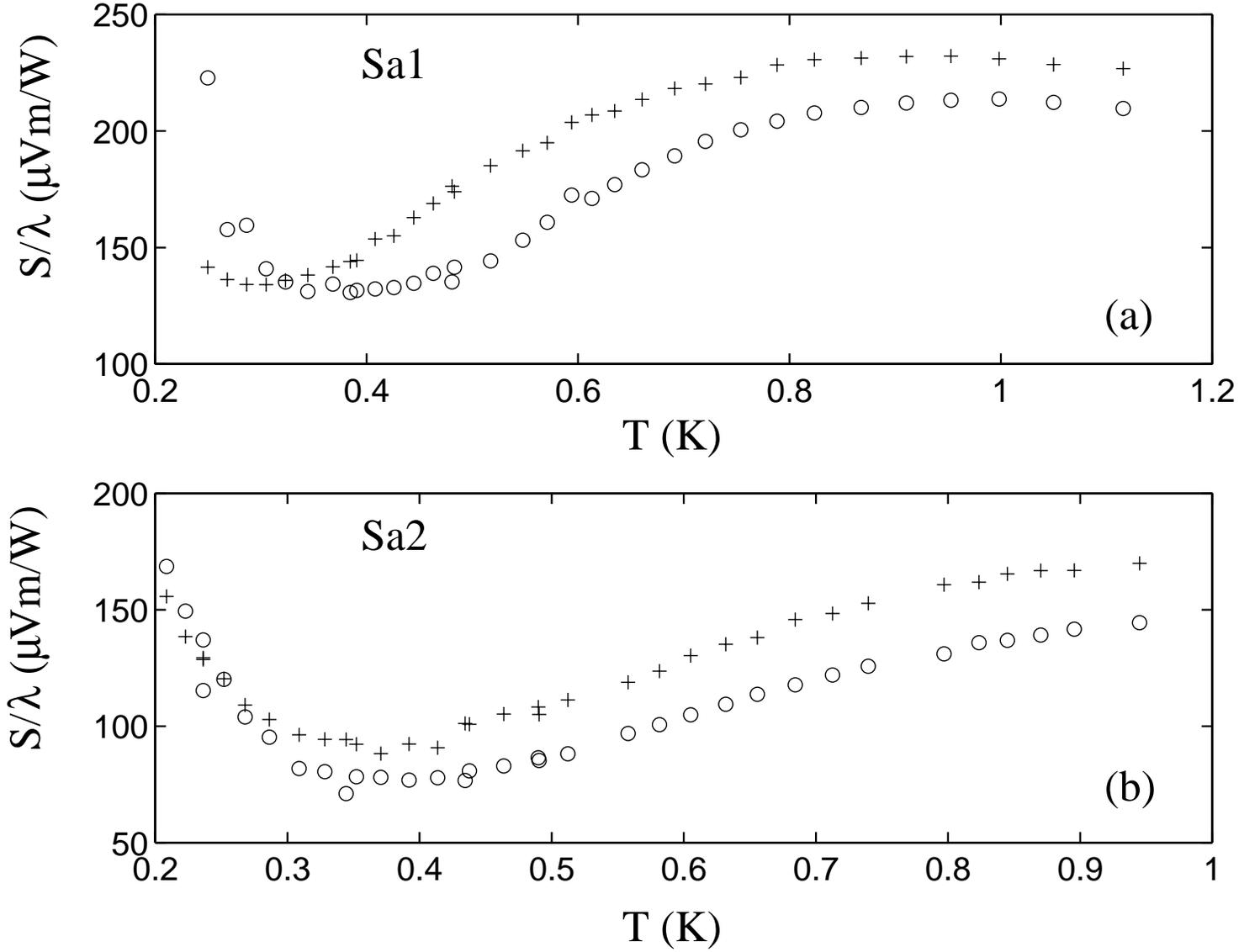}
\caption{The ratio of
the thermopower $S$ over the thermal conductivity $\lambda$ as a
function of T at: a) $P_{s}=0.93\times 10^{15}$m$^{-2}$ and b)
$P_{s}=1.25\times 10^{15}$m$^{-2}$. The circles are the experimental
data and the crosses are the theoretical values.}
\end{figure}
\end{document}